\documentclass[aps,prb,twocolumn,floatfix,showpacs]{revtex4-1}
\usepackage{epsf}
\usepackage{epsfig}
\usepackage{color}
\usepackage{graphicx}
\usepackage{amssymb}
\usepackage{amsmath}
\usepackage{color}
\usepackage{float}
\usepackage{comment}

\def\ep{{\epsilon}}

\def\k{{{\bf k}}}
\def\om{{\omega}}

\def\nnu{{\nonumber}}
\def\CPA{\scriptscriptstyle CPA}
\def\scpa{{\Sigma^{\scriptscriptstyle CPA}_{c/f}}}
\def\zcpa{{Z^{\scriptscriptstyle CPA}_{c/f}}}
\def\zccpa{{Z^{\scriptscriptstyle CPA}_{c}}}
\def\zfcpa{{Z^{\scriptscriptstyle CPA}_{f}}}

\def\g{{\bf{g}}}

\def\beq{\begin{equation}}
\def\eeq{\end{equation}}
\def\beqa{\begin{eqnarray}}
\def\eeqa{\end{eqnarray}}

\def\g0{{\gamma_0}}

\def\Im{{\mbox{Im}}}

\def\prb{Phys.\ Rev.\ B }
\def\prl{Phys.\ Rev.\ Lett.\ }

\def\JPCM{J.\ Phys.\ Condens.\ Matter }

\def\RMP{Rev.\ Mod.\ Phys.\ }

\def\EPL{ Europhys.\  Lett.\ }

\newcommand\bear{\begin{eqnarray}}
\newcommand\eear{\end{eqnarray}}
\newcommand\bea{\begin{align}}
\newcommand\ena{\end{align}}
\begin{document}

\title{Non-Fermi liquid behavior from dynamical effects of impurity scattering in correlated Fermi liquids}
\author{N.\ S.\ Vidhyadhiraja and Pramod Kumar}
\affiliation{Theoretical Sciences Unit\\Jawaharlal Nehru Centre for Advanced Scientific Research,\\
 Jakkur, Bangalore 560 064, India.}

\begin{abstract}
              The interplay of disorder and interactions is a subject of perennial interest. In this work, we have investigated the effect of disorder due to chemical substitution
 on the dynamics and transport properties of correlated Fermi liquids. 
A low frequency analysis in the concentrated and dilute limits shows that the dynamical local potentials arising through disorder averaging generate a linear (in frequency) term
 in the scattering rate. Such non-Fermi liquid behavior (nFL) is investigated in detail for Kondo hole substitution in heavy fermions 
within dynamical mean field theory. We find closed form expressions for
the dependence of the static and linear terms in the scattering rate on substitutional 
disorder and
model parameters. We argue that the low temperature resistivity will acquire a linear
in temperature term, and show that the Drude peak structure in the optical conductivity will
disappear beyond a certain disorder $p_c$, that marks the crossover from lattice
coherent to single-impurity behavior.  A full numerical solution of
the dynamical mean field theory equations reveals that the nFL term will show up significantly only in certain regimes, although it is present for any non-zero disorder concentration in principle. 
We highlight the dramatic changes that occur in the quasiparticle scattering rate
in the proximity of $p_c$. Remarkably, we find that the nFL behavior due to dynamical effects of impurity scattering  has features that are distinct from those arising through Griffiths singularities or distribution of Kondo scales. Relevance of our findings to experiments on alloyed correlated systems is pointed out.   
\end{abstract}


\maketitle

\section{Introduction}

There exist a number of metallic doped systems for which the Fermi
liquid (FL) theory is clearly violated in transport and thermodynamic properties~\cite{stew,maple}. Most theoretical scenarios for non-Fermi liquid (nFL) 
behavior require the proximity of some kind of singularity in the phase diagram, such as a quantum critical point (QCP)~\cite{colm,rosch,vidh00}, Griffiths 
singularities (GS) close to magnetic instabilities~\cite{vojta} or van Hove singularities~\cite{sebastian}. Other proposals for nFL include distribution of Kondo scales
~\cite{mira} and multichannel Kondo effect ~\cite{chang}. The nFL behavior in a wide range of materials has been explained through these proposals. 

Nevertheless, there is an expanding set of  correlated 
systems for which the nFL behavior does not fit the existing scenarios~\cite{stew,maple,maple1,varma}.
For example, in a recent study of the heavy fermion material Ce$_{1-p}$La$_p$B$_6$~\cite{naka}, a resistivity
of the form $\rho(T)=\rho_0 + AT^\alpha$ was found for $p\gtrsim 0.4$, where the non-universal and fractional
exponent $\alpha$, was found to be $p$-dependent. In the FL regime ($p\lesssim 0.4$), the coefficient of the
 quadratic term of the resistivity apparently diverges at the crossover from FL to nFL regime. The optical scattering rate of several U-based materials has been found to have
a linear-in-frequency term~\cite{degiorgi}. A consistent theoretical explanation for such unusual nFL behavior has not yet been found. 

The $T=0$ scattering rate in clean Fermi liquids has the form $\Gamma_{sc}(\om) \stackrel{\om\rightarrow 0}{
\longrightarrow} a_2\om^2$,
The static effects of scattering by quenched substitutional disorder in Fermi-liquids contribute a constant term $a_0$ to $\Gamma_{sc}(\om)$~\cite{mutou,grenz}. 
In this work, we show that dynamical effects of impurity scattering can give rise to the type of aforementioned unusual nFL behavior.  
The $\Gamma_{sc}(\om)$ is shown to acquire an additional
linear in frequency term, $a_1\om$. We also demonstrate that an excellent power-law description of $\Gamma_{sc}(\om)$ is possible over two decades
in $\om$, that yields a doping dependent exponent. 

 In heavy fermion systems such as CeCu$_6$, the substitution of Ce by La is of the
Kondo hole (KH) type while that of Cu with Au  is referred to as ligand-field type~\cite{grenz}.
In this work, we will explicitly consider the KH type of substitution where the
orbital energy of the $f$-level of the substituted non-magnetic atom is high enough that it
 decouples from the conduction band. Any type of random substitution
breaks translational invariance, and leads to site and bond disorder. Random chemical substitution of the type $A_{1-p}B_p$ is usually incorporated through a
probability distribution of the parameters in the Hamiltonian.  Although we have explicitly
considered site disorder with a binary distribution, our arguments will be shown to be applicable 
to other types of disorder and distributions.

\section{Model and formalism}

The Hamiltonian for the periodic Anderson model (PAM), which is appropriate for heavy fermion systems,
may be expressed in standard second-quantized notation
as 
\beq
H_{PAM}=-\sum_{\langle ij\rangle\sigma} t_{ij} \left( c^\dagger_{i\sigma} c^{\phantom{\dagger}}_{j\sigma} + {\rm h.c}\right)  + \sum_{i} H_{ii}  
\eeq
where $t_{ij}$ represents nearest neighbour hopping and $c_{i\sigma}$ is the conduction (c)-electron annihilation operator for site $i$ and spin $\sigma$.
Within dynamical mean field theory (DMFT)~\cite{geor,prus}, which is exact in the limit of infinite dimensions $D$, the hopping
$t_{ij}$ must be rescaled as $t_{ij}\propto t_*/\sqrt{D}$.  The site-diagonal part, given by 
$H_{ii}=\ep_{c} \sum_\sigma c^\dagger_{i\sigma} c^{\phantom{\dagger}}_{i\sigma} + \ep_{f} \sum_\sigma f^\dagger_{i\sigma} f^{\phantom{\dagger}}_{i\sigma}+
V(\sum_\sigma f^\dagger_{i\sigma} c^{\phantom{\dagger}}_{i\sigma} + {\rm h.c}) + Un_{fi\uparrow} n_{fi\downarrow}$ represents
the orbital energies, hybridization between $c$ and $f$ electrons and the cost of double occupancy of the $f$-orbital respectively.

The coherent potential approximation (CPA) is the best single-site approach to study
the interplay of  
disorder with interactions in strongly correlated systems~\cite{grenz,pott}. Even though the CPA ignores inter-site coherence and coherent
 back scattering effects (and hence Anderson localization), the effects of configurational averaging
are well accounted.  A dynamical CPA within DMFT can be formulated that takes into account the dynamical effects of impurity scattering~\cite{mutou,grenz}. In this approach,
the arithmetically averaged local c-electron Green's function called $G^{\CPA}_c$ is given
by
\beq
G^{\CPA}_c(\om)=\frac{1-p}{\om^+ - \ep_c - \Sigma_c - S(\om)} + \frac{p}
{\om^+ - \ep_c - S(\om)}
\label{eq:gccpa}
\eeq
where the $c$-self energy is $\Sigma_c(\om)=V^2 (\om^+ -\ep_f -\Sigma_f(\om))^{-1}$ and $\Sigma_f(\om)$ is the $f$-self-energy. 
The dopant concentration is denoted by $p$, and $S(\om)$ is Feenberg self-energy~\cite{econ} that represents the hybridization with the host.
In the clean case ($p=0$), the $S(\om)$ is a functional of the local $c$-Green's function~\cite{pram}. 
In the disordered case ($p\in (0,1]$), the sum over all neighbouring sites in the Feenberg renormalized perturbation series~\cite{econ} can be reduced, within DMFT, 
to a probabilistic averaging of the magnetic and Kondo hole Green's functions. Thus $S(\om)$ becomes
a functional of the averaged (CPA) Green's function, 
$S(\om)=S[G^{\CPA}_c]$. It is given by the condition that
the CPA restores translational invariance, hence  $G^{\CPA}_c$ is given by
\beqa
G^{\CPA}_c(\om)= \frac{1}{N}\sum_\k \frac{1}{\om^+ - \ep_c - \ep_\k - \Sigma_c^{\CPA}} \nnu \\
= {\cal{H}}\left[\gamma(\om)\right] 
  = \frac{1}{\gamma(\om) - S(\om)}
\label{eq:hilb}
\eeqa
where $\gamma=\om^+ - \ep_c - \Sigma_c^{\CPA}$ and ${\cal{H}}[z]=\int^\infty_{-\infty}d\ep\,\rho_0(\ep)\,(z-\ep)^{-1}$ is the Hilbert
transform of $z$ with respect to the non-interacting density of states, $\rho_0(\ep)$.
 For the semi-elliptical density of states, $\rho_0(\ep)=\sqrt{1-\ep^2/t_*^2}/(2\pi t_*)$, corresponding to the Bethe lattice, the Feenberg self-energy is simply given by 
$S(\om)=t_*^2G^{\CPA}_c(\om)/4$. This allows us to relate the CPA self-energy
directly to the $\Sigma_f$ as shown below.

As expressed by Eq.\ (\ref{eq:gccpa}), the CPA Green's Function is given by
\begin{eqnarray}
G^{\CPA}_c(\omega)=\frac{1}{\Gamma-\Sigma^{\CPA}_c}=\frac{1-p}{\Gamma -\Sigma_c}+\frac{p}{\Gamma}\,,
\label{eq1}
\end{eqnarray}
where $\Gamma=\omega^+-\epsilon_c-S(\omega)$.
From Eq.\ (\ref{eq1}), the CPA self-energy can be extracted as
\begin{equation}
\Sigma^{\CPA}_c=\frac{\Gamma \Sigma_c(1-p)}{\Gamma - p\Sigma_c} \,.
 \label{eq2}
\end{equation}
The above can be inverted to get $\Gamma$ in terms of $p$, $\Sigma_c$ and $\Sigma^{\CPA}_c$ as
\begin{equation}
\Gamma=\frac{p\Sigma_c\Sigma^{\CPA}_c}{\Sigma^{\CPA}_c-\Sigma_c(1-p)} \label{eq3}
\end{equation}
For a Bethe lattice,
\begin{eqnarray}
S(\omega)=\frac{t_{*}^2}{4}G^{\CPA}(\omega)=\frac{t_{*}^2}{4}\frac{1}{\Gamma - \Sigma^{\CPA}_c} \nnu\label{eq4}\,,
\end{eqnarray}
and for $t_{*}=2$
\begin{equation}
S(\omega)=\frac{1}{\Gamma -\Sigma^{\CPA}_c}  \label{eq5}\,.
\end{equation}
Substituting $S(\om)=\om^{+}-\epsilon_c-\Gamma$ in Eq.\ (\ref{eq5}), we get
\begin{equation}
\omega^+ -\epsilon_c -\Gamma=\frac{1}{\Gamma-\Sigma^{\CPA}_c}\,.
\label{eq6}
\end{equation}
In Eq.\ (\ref{eq6}), if we substitute for $\Gamma$, using Eq.\ (\ref{eq3}), we will get an equation relating $\Sigma^{\CPA}_c$ to $\Sigma_c$. 
Alternatively if we substitute for $\Sigma^{\CPA}_c$, using Eq.\ (\ref{eq2}), we will get an equation for $\Gamma$ in terms of $\Sigma_c$. 
In Eq.\ (\ref{eq6}), substituting $\Gamma$ using (\ref{eq3}), we get
\begin{align}
\Sigma^{\CPA}_c\left(\Sigma_c-\Sigma^{\CPA}_c\right)\Big[(\omega^+ -\epsilon_c)(\Sigma^{\CPA}_c-\Sigma_c(1-p))\nnu \\
-p\Sigma_c\Sigma^{\CPA}_c\Big]-
\left(\Sigma^{\CPA}_c-\Sigma_c(1-p)\right)^2=0 
\label{eq:cubic}
\end{align}
This is a cubic equation for $\Sigma^{\CPA}_c$. Given the local self-energy
$\Sigma_c(\om)$, this equation may be used to obtain the CPA self-energy
for any $p$. We study this equation in detail in the next section.

\section{Low frequency Analysis}
\label{sec:lfa}

We have carried out a low frequency analysis of Eq.\ (\ref{eq:cubic}) in the concentrated ($p \rightarrow 0$)
and dilute limits ($p\rightarrow 1$). Our basic premise is that the $f$-moments are completely screened,
and hence have a local FL form of the self-energy, $\Sigma_f(\om)\stackrel{\om\rightarrow 0}{\longrightarrow} \Sigma_f(0) +
\om(1-1/Z) - iA^\prime\om^2$. Using this, the $c$-self-energy $\Sigma_c(\om)=V^2[\om^+ - \ep_f -\Sigma_f(\om)]^{-1}$ can be Taylor expanded as
\beq
\Sigma_c(\om)=\frac{V^2}{\om -\ep_f -\Sigma_f(\om)} \stackrel{\om\rightarrow 0}{
\longrightarrow} -\frac{V^2}{\ep_f^*}\left(1+\frac{\om}{Z\ep_f^*} + i\frac{A^\prime}{\ep_f^*}\om^2\right)
\label{eq:cse}
\eeq
where the renormalized $f$-level, $\ep_f^*=\ep_f + \Sigma_f(0) \in {\cal{R}}\; \forall p$.
The above may be used in Eq.\ (\ref{eq:cubic}) to find the low frequency form
for $\Sigma^{\CPA}_c(\om)$.  

Before delving into the details of the calculations, we summarize our main result.
We find that 
\beq
\Sigma^{\CPA}_c(\om) \stackrel{\om\rightarrow 0}{\longrightarrow} S_0 + S_1\om + S_2\om^2 + {\cal{O}}(\om^3)
\label{eq:quadcpa}
\eeq
where $\{S_i\}$ have finite imaginary parts that depend on $p$.  
The above expression, which shows
that the CPA self-energy has a distinctly nFL form, embodied in the linear in $\om$ term
in the imaginary part, is the central result of our work. A finite
linear term in the ${\rm Im}\Sigma^{\CPA}_c$ in addition to the well-known residual ($\om=0$) and quadratic terms
has broad consequences. First, since $\Sigma^{\CPA}_c = V^2[\om^+ - \ep_f - \Sigma_f^{\CPA}]^{-1}$,
 the CPA self-energy of the $f$-electrons will also have an nFL form, even though the local self-energy has
a standard FL form.  It is easy to show using Eq.\ (\ref{eq:gccpa}) that ~\cite{grenz} 
\beq
{\rm Im}\Sigma_f^{\CPA}(\om)=\frac{1}{1-p}{\rm Im}\Sigma_f(\om) + \frac{p}{1-p}{\rm Im}\frac{V^2}{\om^+ - \ep_c - S(\om)}\,.
\eeq
The above relation implies that the nFL part in the CPA self-energy arises through the contribution
from the imaginary part of the self-consistently determined dynamical hybridization or the local potential.
Second, the scattering rate, $\Gamma_{sc}(\om)\propto -{\rm Im} \Sigma^{\CPA}_c$  will also have a linear non-Fermi
liquid type term in addition to the static potential scattering and  the quadratic electron-electron scattering terms.  
As an inevitable consequence, transport quantities would display an nFL form. For example, the resistivity would have the 
low temperature form $\rho(T) =\rho(0)+ A_RT + B_RT^2$, for a general particle-hole asymmetric case, as shown later (section~\ref{transp}).
Thermodynamics quantities like the specific heat ($C$) will also be affected.
A linear term in the imaginary part of the self-energy naturally leads to a $\om\ln\om$ term in the 
real part. From the expression of specific heat~\cite{vidh07}, it is easy to see that 
a $\ln T$ contribution would arise in $C/T$.

A straightforward generalization of  Eq.\ (\ref{eq:gccpa}) to the case of ligand-field substitution 
 may be carried out~\cite{grenz}. A low frequency analysis similar to the one done above for
KH disorder shows that the linear term would arise even for this case. Similarly,
substitutional disorder in the Hubbard model~\cite{laad} will also yield similar results, since the 
dynamical CPA equations for the local Green's function are exactly the same as Eqs.\ (\ref{eq:gccpa}) and (\ref{eq:hilb})
with the $\Sigma_c$ being replaced by the local self-energy of the interacting electrons.
The CPA self-energy will thus have contributions
from the FL self-energy and the dynamical local potentials which will again lead to a linear (in frequency) term in the scattering rate.
 Thus our findings have implication for transition metal oxides and other systems
for which the Hubbard model is appropriate. Here, we have considered a binary distribution of site energies. A generalization
to other discrete or continuous distributions can be made simply by generalizing 
Eq.\ (\ref{eq:gccpa}) to a general distribution, ${\cal{P}}(\ep_{ci})$, in the following way:
\beq
G^{\CPA}_c(\om)=\int^\infty_{-\infty} \,d\ep_{ci}\,{\cal{P}}(\ep_{ci})\frac{1}
{\om^+ - \ep_{ci} - \Sigma_c - S(\om)}\,.
\eeq
The rest of the analysis proceeds in exactly the same way as for the binary distribution, and hence this will also yield similar nFL behavior. 
Next, we find closed-form approximate expressions for the coefficients of the
scattering rate (Eq.\ (\ref{eq:quadcpa}))
in the dilute and concentrated limits.

\subsection{Analytical expressions for static and linear terms in scattering rate}

We expect to find analytic solutions for the CPA self-energy, hence expanding $\Sigma^{\CPA}_c(\om)=S_0+S_1\omega$, and using  Eq.\ (\ref{eq:cse}) 
in Eq.\ (\ref{eq:cubic}) with $\left(\omega^+ -\epsilon_c \rightarrow -\epsilon_c\right) \ \text{for} \ \om \rightarrow 0$, we get
\begin{align}
&\left(S_0+S_1\omega\right)\left(\Sigma_{c0}(1+A\omega) -S_0-S_1\omega\right)\Big[(-\epsilon_c) \nnu \\
 &\left(S_0+S_1\omega-\Sigma_{c0}(1+A\omega)(1-p)\right) 
-p\Sigma_{c0}\left(1+A\omega\right)\nnu \\
&\left(S_0+S_1\omega \right)\Big]-\Big[S_0+S_1\omega-\Sigma_{c0}(1+A\omega)\Big]^2=0\,.
\label{eq:stdn}
\end{align}
where $A=1/(Z\ep_f^*)$.
Substituting $\omega=0$ in the above will yield an equation for the static contribution, namely $\Sigma^{\CPA}_c(0)$. The coefficient of the linear in $\om$ term
may be found by collecting the linear terms. This will be done in the following subsections in the concentrated ($p\rightarrow 0$) and dilute ($p\rightarrow 1$) limits. 

\subsubsection{Concentrated limit $p\rightarrow 0$}

After a lengthy and tedious, but straightforward calculation, we find the 
explicit dependence of $\{S_i\}$ (Eq.\ (\ref{eq:quadcpa})) on $p$  
for a symmetric conduction band ($\ep_c=0$). In the concentrated limit, $p\rightarrow 0$, the static part of $\Sigma^{\CPA}_c$
 is given by 

\begin{align}
{\rm Re}(S_0)=&(1-p)\Sigma_{c0} + \frac{p(1-p)^2}{2}\Sigma_{c0}^3
\nnu \\
{\rm Im}(S_0)=& -p(1-p)\Sigma_{c0}^2 \times \left[ 1- \frac{(1-p)^2}{4}\Sigma_{c0}^2\right]^{1/2}\,
\label{eq:static}
 \end{align}

 where $\Sigma_{c0}=-V^2/\ep_f^*$. 
The ${\rm Im} (S_0)$ leads to a finite $T=0$ residual resistivity.
The coefficient of the linear term, ${\rm Im} (S_1)$ is given by 

\beq
{\rm Im}(S_1)=\frac{p}{Z\ep_f^*}{\rm Im}\left[\frac{(\delta\Sigma_{c0})^2(\Sigma_{c0}-\delta\Sigma_{c0})}
{(\delta\Sigma_{c0})^2 - 2(\delta\Sigma_{c0} -\Sigma_{c0})^2}\right]
\label{eq:ima}
\eeq

where $\delta\Sigma_{c0}=-\frac{(1-p)\Sigma_{c0}^2}{2}\left[(1-p)\Sigma_{c0} + 2i\right]$ is an ${\cal{O}}(1)$ number in the limit $p\rightarrow 0$.
Thus, the nFL contribution is seen to be proportional to $p$ at low dopant concentrations.
It is also important to note that the quasiparticle weight (QpW) $Z$  does not appear in the static 
part (Eq.\ (\ref{eq:static})), but does appear in the dynamics, and arises purely because
 of the linear term in the FL form of the local self-energy, $\Sigma_f(\om)$. Hence, the dynamical
 effects of potential scattering are responsible for nFL behavior due to substitutional doping.

\subsubsection{Dilute limit: $p\rightarrow 1$}

In the dilute limit ($p\rightarrow 1$), the hybridization is determined through the
non-interacting density of states, $\rho_0(\ep)$. Hence we get 
\begin{equation}
\Sigma^{\CPA}_c(\om)=\frac{V^2(1-p)}{\omega^+ -\epsilon_f-\Sigma_f(\omega)-p\Delta_0(\om)}
\end{equation}
where ${\rm Im} \Delta_0(\om)= -\pi V^2\rho_0(\om-\ep_c)$. In the strong coupling limit,
and for low frequencies $\om \lesssim \om_L=ZV^2/t_*$, the $\rho_0(\om)$ may be taken to be a constant,
 yielding the low frequency form of the CPA self-energy as:
\begin{equation}
\Sigma^{\CPA}_c \stackrel{\om\rightarrow 0}{\longrightarrow} \frac{V^2(1-p)}{\frac{\omega}{Z}-{\tilde{\epsilon}}^*_f + ip{\bar{\Delta}}_0+iA^\prime\om^2}
\label{eq:siam}
\end{equation}

where ${\tilde{\epsilon}}_f^*=\epsilon_f+\Sigma_f(0)+p {\rm Re} \Delta(0)$, and ${\bar{\Delta}}_0
=\pi V^2\rho_0(-\ep_c)$. From the above expression, it is easy to see that in the dilute limit, all the three coefficients of
Eq.\ (\ref{eq:quadcpa}) (divided by the number of magnetic atoms $(1-p)$) obtained through a Taylor expansion of Eq.\ (\ref{eq:siam}) 
around $\om=0$  will remain non-zero. Thus for $p\rightarrow 1$, if either the effective $f$-level or the conduction band centre ($\ep_c$)
are non-zero, a linear term can be obtained in the temperature dependence of the resistivity, as discussed later in section~\ref{sec:results}.

\subsection{General considerations for the quasiparticle weight}
\label{sec:qpw}

Before we discuss transport, we will briefly
consider the behavior of the QpWs as derived through the CPA quantities.
This is important, because much of the low frequency and low temperature physics in clean
systems can be gleaned through the renormalized non-interacting limit, where in the
bare parameters such as the hybridization, $V^2$ or the bandwidth, $t_*$ are renormalized 
by the QpW. As an example of the consequence of such renormalization, the integrated
spectral weight contained in the Drude peak of the optical conductivity is proportional to $Z$~\cite{davjpcm} in the clean system.
In the disordered case, since the response functions are determined by the CPA self-energies, $\Sigma^{\CPA}_{c/f}$,
rather than the local self-energies, a clear picture must be obtained of the $\zcpa$ 
defined as
\beqa
\frac{1}{\zcpa}&=&\left(1-\frac{\partial  \Sigma^{\CPA}_{c/f}(\om)}{\partial\om}|_{\om\rightarrow 0}\right) \;\;\;{\rm and} \\
\frac{1}{Z^{\CPA}_{c/f\,R}}&=&{\rm Re}\left(\frac{1}{\zcpa}\right)
\label{eq:zcpa}
\eeqa
From Eqs.\ (\ref{eq:quadcpa}) and (\ref{eq:ima}) above,
it is clear that the QpWs defined above through the CPA self-energy would
in general be complex. Furthermore, unlike the clean case, where the QpW for the 
c-electrons is proportional to that of the f-electrons, the $\zccpa$ and $\zfcpa$
may behave entirely differently, because of the finite imaginary parts of either
of these. This may be seen as follows. The CPA self-energy of the $c$-electrons
is related to that of the $f$-electrons through
\beq
\Sigma^{\CPA}_{c}=\frac{V^2}
{\om^+ -\ep_f - \Sigma^{\CPA}_{f}}
\stackrel{\om\rightarrow 0}{\longrightarrow} \frac{V^2}
{\om/\zfcpa -\ep_f^*}\,
\eeq
where $\ep_f^*$ is a complex number. This implies the following for the corresponding
QpWs:
\beq
\frac{1}{\zccpa}=1+\frac{V^2}{(\ep_f^*)^2}\frac{1}{\zfcpa}
\eeq
which implies that the real part of the CPA $c$-electron QpW is dependent on
both, the real and imaginary, parts of the $\zfcpa$. 

\section{Transport: Analytical considerations}
\label{transp}

In this section, we will discuss the consequences of an  
anomalous CPA self-energy (Eq.\ (\ref{eq:quadcpa})) on transport quantities, namely
resistivity and optical conductivity. 

\subsection{Resistivity}

For the disordered PAM on a hypercubic lattice, the dc conductivity (within CPA) is given
 by~\cite{pram}
\beqa
&\sigma_{dc}(T)=\sigma_0\int^{\infty}_{-\infty}\,d\om \left(-\frac{\partial n_F(\om)}{\partial \om}\right) \tau_{dc}(\om) 
\label{eq:dcc}\\
{\mbox{where}}\; &\tau_{dc}(\om)=\frac{\pi D^{\CPA}_c(\om)}{\gamma_I(\om)} + 2\left(1-\gamma(\om)\right)
G^{\CPA}_c(\om)\,, \label{eq:taudc}
\eeqa
and $\gamma(\om)$ is defined below Eq.\ (\ref{eq:gccpa}), $\gamma_I= {\rm Im} \gamma=-{\rm Im} \Sigma^{\CPA}_c$
and $n_F(\om)=\left(\exp(-\om/T)+1\right)^{-1}$ is the Fermi-Dirac distribution function
(with $T$ being the temperature). 
The local self-energy maintains
a FL form in $\om$ {\em and} $T$ for all $p$, so even though the $\om$ dependence
of $\scpa$ is anomalous, the explicit $T$ dependence of $\scpa$ will remain FL-like
for all $p$.
As an important consequence, since the $\tau_{dc}(\om;T)$ depends on ${\rm Im} \scpa$ (Eq. (\ref{eq:taudc})), the frequency dependence of the scattering rate will be anomalous, but the temperature dependence will remain FL-like ($\sim T^2$). Such a 
$\tau_{dc}(\om;T)$ when substituted in Eq.\ (\ref{eq:dcc}) will nevertheless yield a linear in temperature term in the dc conductivity as shown below.

We will revisit the clean case first, and then consider the disordered case. For $p=0$,
the $\tau_{dc}(\om)$ of a clean FL may be approximated
at the lowest $(\om,T)$ by using the FL expansion of the local self-energy as
\beq
\tau_{dc}^{FL}\simeq \pi\frac{\pi D_c(0)}{a\om^2+bT^2}
\eeq
which is just a symmetric Lorentzian centred at $\om=0$, and a width $\sim T$.
For an {\em even function} $\tau_{dc}(\om)$, the integral in Eq.\ (\ref{eq:dcc}) will yield a
conductivity that is an {\em even} function of temperature, because the derivative of the Fermi function
is an even function of $\om/T$. Thus the resistivity will have the form $\rho_{dc}(T)=AT^2$.

In the presence of disorder, if only the static contribution to  $\Sigma^{\CPA}_c$ is included,
then the $\tau_{dc}(\om)$ remains a symmetric Lorentzian, hence the conductivity will again be an even function
of $T$, thus the resistivity will acquire the form $\rho_{dc}(T)=\rho_0 + AT^2$. 

The nFL contribution embodied in the linear in $T$ term arises from the dynamical, 
linear in $\om$ contribution to
the imaginary part of the CPA self-energy, as argued below.
We have shown that the $\Sigma^{\CPA}_c$  acquires an nFL form
for $p\neq 0$, where due to dynamical effects of impurity scattering, a linear in $\om$ term arises (Eq.\ (\ref{eq:quadcpa})). 
Since $\gamma_I={\rm Im}(\om^+-\ep_c-\Sigma^{CPA}_c(\om))$, the 
$\tau_{dc}(\om)$ (equation~(\ref{eq:taudc})) is no longer an even function; in fact, it becomes a shifted Lorentzian 
of the form
\beq
\tau_{dc}^{nFL}\simeq \pi\frac{\pi D^{\CPA}_c(0)}{a_0+a_1\om + a_2\om^2+bT^2}\,.
\eeq
Now, for a strongly asymmetric $\tau_{dc}(\om)$ (about $\om=0$), the conductivity will pick up
linear in $T$ terms. As a simple illustration,
a square pulse form of $\tau_{dc}(\om)=\theta(-\om)\theta(\om+|a|)$ yields $\sigma(T)=\sigma_0(1/2+T/|a|)$,
which is clearly a non-Fermi liquid form. Thus, we have argued that an nFL frequency dependence of the
CPA self-energy will yield
a nFL temperature dependence of dc conductivity.

\subsection{Optical Conductivity}

In this section, we will explore the consequences of Kondo hole substitution on
the low frequency features of the optical conductivity.
The optical conductivity at zero temperature within DMFT is given by~\cite{pram} 
\begin{align}
\sigma(\om)=\frac{\sigma_0}{\omega}
\int_{-\infty}^0d\om^{\prime}\int_{-\infty}^{\infty}d\epsilon \rho_0(\ep)D(\om^{\prime},\epsilon)D(\om+\om^{\prime},\ep)
\end{align} 
where
\begin{equation}
D(\om,\ep)=-\frac{1}{\pi}\Im\frac{1}{\gamma(\om) -\ep}
=\frac{\gamma_I/\pi}{(\gamma_R -\ep)^2+\gamma_I} \nnu\,
\end{equation} 
$\gamma(\om)=\gamma_R+i\gamma_I$ and $\gamma_I\geqslant 0$. 
Thus,
\begin{align}
\sigma(\omega)=&\frac{\sigma_0}{\pi^2\om}\int_{-\infty}^0d\om^{\prime}\gamma_I(\om^\prime)
\gamma_I(\om+\om^\prime)\,\times\nnu \\
&\int_{-\infty}^{\infty}\frac{d\ep\rho_0(\ep)}{\left[(a_1-\epsilon)^2+b_1^2\right]\left[(a_2-\epsilon)^2+b_2^2\right]}
\end{align}
where $a_1=\gamma_R(\om)$, $b_1=\gamma_I(\om)$, $a_2=\gamma_R(\om+\om^\prime)$ and $b_2=\gamma_I(\om+\om^\prime)$.
Assuming a wide, flat band for simplicity,
$\rho_0(\ep)=\frac{1}{2W}\theta(W-|\ep|)$,
we can carry out the integral over $\ep$ exactly to find
\begin{equation}
\sigma(\omega)=\frac{\sigma_0}{2\pi^2\om W}\int_{-\om}^{0}\,d\om^\prime\frac{\pi Q}{2\sqrt{b_1 b_2}}(B+\frac{1}{B})\label{eq:q}
\end{equation}
where $A=\frac{a_1-a_2}{2\sqrt{b_1 \ b_2}}$ and $B^2=\frac{b_1}{b_2}$.
In the limit of $\om\rightarrow 0$, the integrand may be Taylor expanded 
about the zero frequency limit using the Taylor expansion of the CPA self-energies.
This finally yields the following expression for the zero temperature
optical conductivity:
\begin{equation}
\sigma(\om)=\frac{\sigma_0}{4 \pi W \bar{\gamma}_I}\frac{1}{(\om / 2\bar{\gamma}_IZ^{\CPA}_{cR})^2+1}
\label{eq:fopt}
\end{equation}
 
where $\bar{\gamma}_I=\gamma_I(0)$ is just the static part of the CPA self-energy.

We see from the above equation that the low frequency form of $\sigma(\om,T=0)$ is
a Drude peak that has a Lorentzian shape. In the clean limit $(p=0)$ and at zero
temperature, the
imaginary part of $\Sigma^{CPA}_c(0)$ vanishes, and $Z^{\CPA}_{cR}$ is a finite
positive number. Thus, we recover a Dirac delta functional form of the Drude peak
in the clean limit. For finite $p$,  the $\bar{\gamma}_I$ becomes finite, and hence
the Drude peak broadens into a Lorentzian, with an integrated spectral weight remaining proportional to $Z^{\CPA}_{cR}$.
As will be seen later, the  $Z^{\CPA}_{cR}$ crosses zero and becomes negative beyond
 a certain $p_c$, thus marking the 
complete destruction of the Drude peak, and hence a complete crossover to the 
single-impurity regime. At the present level of approximation in Eq.\ (\ref{eq:q}),
we do not find a explicit contribution of the imaginary part of the QpW, i.e
$Z^{\CPA}_{cI}$, although as shown in section~\ref{sec:qpw}, even the real
part of the $\zccpa$ depends on the real {\em and} imaginary parts of
$\zfcpa$. However, we conjecture that
a higher order Taylor expansion in equations~(\ref{eq:q}) will certainly lead to a manifestation
of the $Z^{\CPA}_{cI}$ and a deviation from the Lorentzian form. 

\section{Results and discussion}
\label{sec:results}
The previous section was entirely focused on getting analytical insights
into the manifestation of the anomalous form of the CPA self-energy in
transport quantities. We were able to analyse the clean and dilute limits for the CPA
self-energy in section~\ref{sec:lfa}, and find the low temperature and low frequency forms of the
dc conductivity and optical conductivity respectively. 
In order to explore the nFL behavior quantitatively in the full range of dopant concentration and at all energy scales,  we have carried out detailed calculations
for the Kondo hole disordered PAM within DMFT. The local moment approach
 (LMA) has been used to solve the effective impurity problem arising within DMFT. 
Within the LMA, which is a diagrammatic perturbation theory based approach, the $f$-self energy is ensured to have a FL form, since adiabatic continuity to the
noninteracting limit is imposed as a constraint. The reader is referred to our earlier work~\cite{pram}
for the detailed implementation of the LMA within DMFT for the clean PAM~\cite{note}.
The coherence peak in the resistivity of clean heavy fermions is found to be at $T\sim\om_L$~\cite{pram}, where $\omega_L=ZV^2/t_*$
is the low energy scale of the local Fermi liquid~\cite{davjpcm}.
 We focus on the frequency region $\om \ll \om_L$, since the nFL behavior is found experimentally at temperatures much below the coherence peak~\cite{stew,maple}.

\begin{figure}[ht]
\centering{
\includegraphics[scale=0.56,clip=]{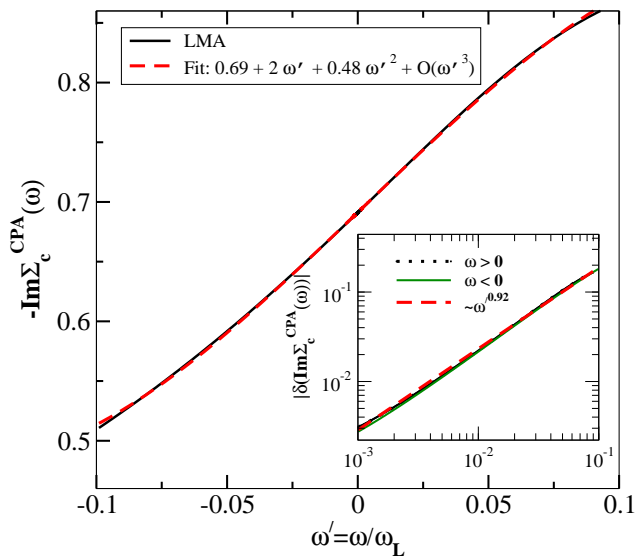}
}
\caption{(color online) In the main panel, the ${\rm Im}\Sigma^{\CPA}_c(\om)$ as a function of scaled frequency $\omega^{\prime}=\omega/\omega_L$ 
has been shown with solid line for $p=0.5$. The dashed line is a polynomial fit (equation~\ref{eq:quadcpa}).
In the inset, the dotted  and solid lines represent $|\delta({\rm Im}\Sigma^{\CPA}_c(\om))| = |{\rm Im}(\Sigma^{\CPA}_c(\om) - S_0)|$ for $\om >0$ and $\om<0$ respectively. The dashed line
 is a power law fit $D(\om^\prime)^\alpha$, that yields a sub-linear 
$\alpha \simeq 0.92$. The model parameters for this calculation are $U=5.30, V^2=0.6, 
\ep_c=0.5$ and $\ep_f\sim -U/2$; the occupancies are
 $n_f\simeq 0.98, n_c\simeq 0.57$ and doping concentration is $p=0.5$.}
\label{fig:plfit}
\end{figure}

In the main panel of Fig.\ (\ref{fig:plfit}), we show (through fitting) that the ${\rm Im}\Sigma^{\CPA}_c(\om)$ does indeed  have the 
 polynomial form of Eq. (\ref{eq:quadcpa}) for $p=0.5$. Additionally, we observe that 
a power law of the form $C+D{\rm sgn}(\om)|\om|^\alpha$ may {\em also}
 be fit  over two decades from $|\om|\sim {\cal{O}}(10^{-3}\om_L)$ upto a certain upper cutoff $\om_c\sim {\cal{O}} (0.1\om_L)$.
This is shown in the inset of figure~\ref{fig:plfit} for both negative and positive frequencies.
Hence the low temperature resistivity,
 as obtained through the scattering rate, is susceptible to a power law interpretation, with the exponent
being a function of disorder (see below). We have verified that the quadratic and 
power law fits are equally good for the entire range of doping.

\begin{figure}[th]
\centering{
\includegraphics[scale=0.56,clip=]{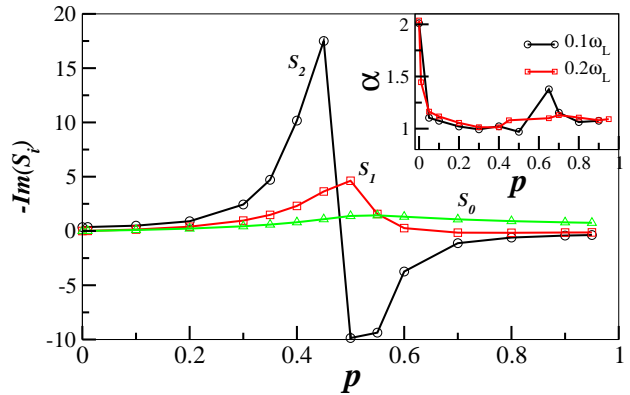}
}
\caption{ (color online)The main panel shows ${\rm Im}(-S_0)$ (triangles), ${\rm Im}(-S_1)$ (squares) and ${\rm Im}(-S_2)$
(circles) per magnetic atom (equation~\ref{eq:quadcpa}). The inset shows
the exponent $\alpha$ as a function of $p$, obtained through a power law fitting of the
low frequency $\Im \Sigma^{\CPA}_c(\om)$ for cutoff equal to $\sim 0.1\om_L$ (circles) and $\sim 0.2\om_L$ (squares). The model parameters for these results are the same
as those for figure~(\ref{fig:plfit}).  }
\label{fig:self_coeff}
\end{figure}
We now consider the dependence of the fitting parameters of the quadratic form
of the CPA self-energy (Eq.\ (\ref{eq:quadcpa})) on $p$. 
The ${\rm Im}(-\{S_i\})$ (Eq.\ (\ref{eq:quadcpa})) per
 magnetic atom, as a function of $p$ are shown in the main panel of Fig.\ (\ref{fig:self_coeff}). 
The first two ($i=0,1$) vanish for  $p\rightarrow 0$,
 but remain finite for all other $p$ including $p\rightarrow 1$. To explain this,
we recapitulate that the coefficients, $S_0$ and $S_1$ develop finite imaginary parts
due to disorder averaging in the CPA, which represents the effects of potential scattering.
For any $p\neq 0$, Kondo holes will be in random positions, thus
impurity scattering will be present, and will lead to the result seen in Fig.\ (\ref{fig:self_coeff}).
The static part
follows a Nordheim rule like behavior, while the linear and quadratic coefficients show an apparent divergence before switching sign
 abruptly at $p\sim 1-n_c\sim 0.43$. Although we see that the linear term is non-zero
over the entire range of $p\in (0,1]$, it becomes significant only in the neighbourhood
of  $p\sim 1-n_c$. We have also seen numerically (as well as from Eqs.\ 
(\ref{eq:ima}) and (\ref{eq:siam})) that for a system which is close to the particle-hole symmetric limit, such nFL behavior would be so weak that it would not show up.
The inset of Fig.\ (\ref{fig:self_coeff}) shows the power law exponent $\alpha$
for two cutoff values, namely $\om_c\sim 0.1\om_L$ (circles) and $0.2\om_L$ (squares). The power law fit depends sensitively
 on the cutoff $\om_c$, which is ambiguous.
 The exponent $\alpha$ is seen to be p-dependent and close to $1$ over a large range of $p$.
Many theoretical studies have pointed out the crossover of collective to single-impurity behavior at $p\sim 1-n_c$~\cite{burdin,kaul,wata}, and
 we see from Fig.\ (\ref{fig:self_coeff}) that indeed dramatic changes could happen in the CPA quantities as $p$ is tuned through this crossover.  

\begin{figure}[ht]
\centering{
\includegraphics[scale=0.6,clip=]{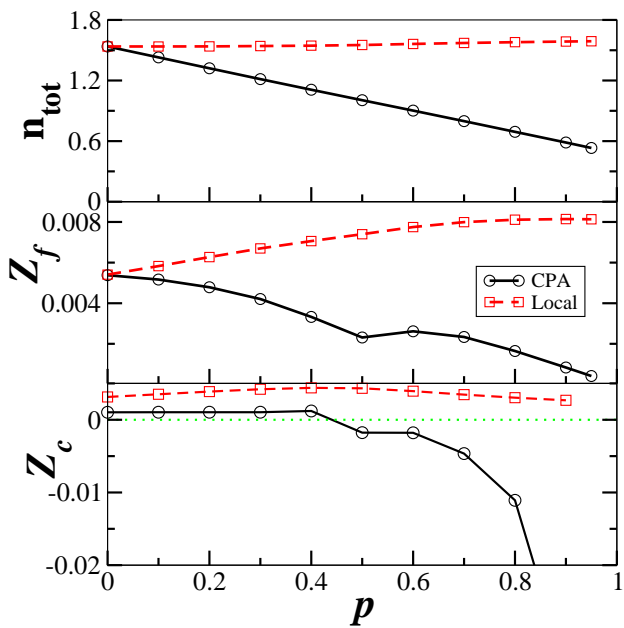}
}
\caption{
(color online)Top panel: The total occupancy, $n_{tot}=n_f+n_c$ as calculated from the local Green's functions (filled squares) and the
 CPA Green's functions (filled circles)
as a function of $p$. Middle Panel: The quasiparticle weight computed through 
the local $f$-self energy (squares) and the real part of the CPA $f$-self-energy (circles) with increasing disorder value $p$. Bottom panel:The quasiparticle weight computed through 
the local $c$-self energy (squares) and the real part of the CPA $c$-self-energy (circles) with increasing disorder value $p$.
The model parameters are same as in Fig.\ (\ref{fig:plfit}). The green dotted line
marks the zero line for the $y$-axis.
}
\label{fig:local_nonlocal}
\end{figure}

Although the CPA quantities display remarkable non-monotonic behavior on varying $p$,
the local quantities are either monotonic with $p$ or remain almost unchanged. 
This is shown in the Fig.\ (\ref{fig:local_nonlocal}), where we compare the occupancy (top panel) and the $f$- and $c$-quasiparticle weights ($Z_f$: middle panel and $Z_c$:
bottom panel respectively) computed through 
local $\Sigma_{c/f}(\om)$ (squares) with the corresponding CPA quantities (circles)
using the real part of the CPA self-energies $\Sigma^{\CPA}_{c/f}(\om)$ . The local quasiparticle weights show a weak dependence on increasing Kondo hole concentration, while the CPA quantities shows non-monotonic behavior.
The dip in $Z^{\CPA}_{fR}$ at $p\sim 1-n_c$ would manifest as a peak in specific heat coefficient or the effective mass. The occupancy, calculated locally, remains almost unchanged, 
while the $n_{tot}^{\CPA}$, given by $n_f+n_c-p$, decreases linearly as expected~\cite{grenz}.  
The $c$-electrons' CPA quasiparticle weight shows dramatic behavior as a function of $p$.
At small $p$, the $Z^{\CPA}_{cR}$ is positive, as expected from continuity to the
clean limit. As $p$ approaches $1-n_c$, the $Z^{\CPA}_{cR}$ rapidly decreases
and becomes negative at $p\gtrsim 1-n_c$. The vanishing of the CPA QpW
will have serious consequences for the optical conductivity. From Eq.\ (\ref{eq:fopt}),
it is clear that the integrated spectral weight contained within the Drude peak will vanish.

\begin{figure}[ht]
\centering{
\includegraphics[scale=0.4,clip=]{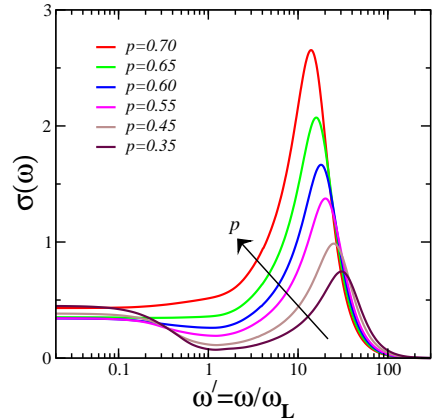}
}
\caption{
(color online) The zero temperature optical conductivity as a function of scaled
frequency $\om^\prime=\om/\om_L$ is shown for various values of doping
$p$. The model parameters for this calculation are $U=5.30, V^2=0.6, 
\ep_c=0.7$ and $\ep_f\sim -U/2$; the occupancies are
 $n_f\simeq 0.97, n_c\simeq 0.43$.
}
\label{fig:optics}
\end{figure}

This is borne out by the $T=0$ optical conductivity results shown in Fig.\ (\ref{fig:optics}). With increasing $p$, the mid-infrared peak ($\sim 50 \om_L$)
narrows and shifts to lower frequencies. As expected above, the low-frequency Drude
peak feature is gradually replaced, with increasing $p$, by a flat featureless line shape beyond $p\gtrsim 1-n_c$.  

The nFL behavior we find is within the framework of CPA. It is important to know
if such behavior is an artifact of CPA, or would it survive beyond CPA. To 
understand this in quantitative detail requires the use of an approach more
 sophisticated than CPA. While such a calculation is beyond the scope of this paper, 
we will nonetheless argue qualitatively that the nFL behavior that we find must manifest
in experimental probes.  The essence of our
 findings is that disorder-averaging
of local Fermi-liquids induces a linear in $\om$ term in the effective self-energy.
Currently, one of the best methods to incorporate local correlations and disorder is the
statistical-DMFT~\cite{mira}. In this method, the local environment of each site is distinct,
and is found self-consistently. If there are no unscreened spins, then this method 
yields all sites to be local Fermi liquids. Nevertheless, since any bulk experimental
 probe such as optical conductivity or dc resistivity, would be determined by disorder-averaged quantities, we conjecture that these quantities might exhibit nFL properties.
In this context, even local optical probes such as infrared microscopy must be considered
macroscopic, since the spatial resolution of such probes is
$\sim 20 nm$~\cite{basov}, which would represent a cluster of hundreds of atoms.
Only a strictly atomic level probe such as scanning tunneling microscopy, would
however be able to distinguish between true nFL behavior arising due to
a breakdown of local Fermi liquid or a disorder-averaging induced nFL behavior.
 A full statistical-DMFT calculation of the local Green's functions and self-energy and
further calculation of response functions must be carried out to verify our conjecture.

\section{Conclusions}
 There have been many recent theoretical studies of substitutional disorder in the Kondo lattice model or the PAM~\cite{grenz,burdin,kaul,wata}. 
It was shown by Kaul and Vojta ~\cite{kaul} that Griffiths singularities (GS) appear
 in a wide range of concentrations leading to nFL behavior. The GS induced nFL has specific `universal' signatures, albeit dependent on a non-universal exponent $\lambda$~\cite{vojta}.
The authors also observe unscreened spins, which would imply that certain sites have  
a vanishing Kondo scale. Such a probability distribution of Kondo scales, where
$P(T_K=0)$ is finite, is also known to yield nFL behavior~\cite{mira}. In a recent work,
a Lifshitz transition~\cite{burdin} is predicted to occur as a function of $p$, which could lead to nFL behavior in the vicinity of the transition.
 The interaction of spin fluctuations with disorder close to a quantum critical point is also known to lead to power law behavior, with a 
disorder-dependent exponent~\cite{rosch}. While inhomogeneities are natural and must be expected in any disordered system,
instabilities such as a QCP and singularities such as GS are necessarily non-generic, i.e they must occur only in specific regions
of the phase diagram. 

While such singularities do give rise to specific nFL behavior, our
work shows that nFL behavior can be quite generic and can arise simply as a consequence of disorder averaging. Hence, 
attributing the deviations from FL to a specific cause in disordered systems needs care. 
 In Ce$_{1-p}$La$_p$B$_6$ for example, we suggest that a
 quadratic polynomial fit must be carried out for all $p$ instead of a partial power law fit. We predict that the fit parameters
 would follow the behavior shown in figure ~\ref{fig:self_coeff}. 
 
 A distinct signature of the nFL behavior we find
 is that it is a macroscopic effect, hence local probes such as scanning tunneling
microscopy should find local Fermi liquid behavior while macroscopic response functions
 would show nFL signatures. However, large area scans would be necessary to rule out
 unscreened spins or Griffiths singularities.

\acknowledgments
We thank H.\ R.\ Krishnamurthy and David E.\ Logan for discussions.
We acknowledge CSIR, India and DST, India for financial support.


\begin{thebibliography}{99}
\bibitem{note} The calculations have been carried out for a hypercubic lattice, for which $\rho_0(\ep)=\exp(-\ep^2/t_*^2)/\sqrt{\pi}t_*$. The unit of energy is $t_*=1$.
\end{thebibliography}
\end{document}